# Compact on-chip power splitter based on topological photonic crystal

PUHUI ZHANG,[1] JIACHENG ZHANG,[1] LINPENG GU,[1] LIANG FANG,[1] YANYAN ZHANG,[2,4] JIANLIN ZHAO,[1] AND XUETAO GAN[1,3,*]

[1]*Key Laboratory of Light Field Manipulation and Information Acquisition, Ministry of Industry and Information Technology, and Shaanxi Key Laboratory of Optical Information Technology, School of Physical Science and Technology, Northwestern Polytechnical University, Xi'an 710129, China*
[2]*School of Artificial Intelligence, OPtics and ElectroNics (iOPEN), Northwestern Polytechnical University, Xi'an 710072, China*
[3]*School of Microelectronics, Northwestern Polytechnical University, Xi'an 710129, China*
[4] *zhangyanyan@nwpu.edu.cn*
[*] *xuetaogan@nwpu.edu.cn*

**Abstract:** We propose and demonstrate an on-chip 1×N power splitter based on topological photonic crystal (TPC) on a monolithic silicon photonic platform. Benefiting from the valley-locked propagation mode at the interface of TPCs with different topological phases, the proposed power splitter has negligible backscattering around the sharp bendings and good robustness to fabrication defects, which therefore enable lower insertion loss, better uniformity, and more compact footprint than the conventional designs. For the fabricated 1×2 (8) power splitter, the uniformity among the output ports is below 0.35 (0.65) dB and the maximum insertion loss is 0.38 (0.58) dB with compact footprint of 5×5 μm$^2$ (10×12 μm$^2$) within a bandwidth of 70 nm. In addition, the topological power splitter only requires simple configurations of TPCs with different topological phases, which is more reliable in design and fabrication compared with the conventional designs.

On-chip power splitter is one of the key components in large-scale photonic integrated circuits (PICs) for splitting and combining the light beam in the applications of signal feedback, monitoring and power distribution. Various structures have been demonstrated to realize power splitters, including directional couplers (DCs) [1,2], multimode interference (MMI) couplers [3,4], plasmonic subwavelength waveguides (SWGs) [5], Y-junctions [6,7], photonic crystal waveguides [2,8,9], and other structures with inversion design [10,11].

The DC based on evanescent-field coupling of two strip waveguides is the most popular method of splitting and combining light in PICs, which promises direct adjustment of the power splitting ratio through proper selection of the coupling lengths. However, it is sensitive to wavelength and fabrication errors, which limits its broadband applications [2]. Compared with DCs, MMI couplers are less wavelength-dependent and offer a wider optical bandwidth. It divides the optical power according to the principle of self-imaging and provides superior performances due to the relative compact footprint and the excellent fabrication tolerance. However, it usually exhibits nonnegligible excess losses. In addition, it has been pointed out that, as the number of output ports increases, the power uniformity and the insertion loss deteriorate while the device footprint increases [3]. The SWGs have been successfully applied to DCs and MMIs to improve their bandwidths as it can engineer optical properties through subwavelength structures, but the gratings are intrinsically sensitive to fabrication errors. Y-junctions have also been a preferred candidate for the power splitting due to their compactness, less wavelength-dependences and low-loss characteristics, but the mode mismatch caused excess loss for the Y-branch domain is relatively high when the branching angle is not sufficiently small. Photonic crystal waveguides are also promising for optical splitters with reduced crosstalk, except for large transmission and bending losses. In addition to these, a series of high-performance power splitters with inversion designs have been proposed. For instance, using a phase gradient algorithm to design a grating metasurface, a power splitter has been realized with arbitrary number of output ports, power ratios and spatial distributions [11]. Note that, as the number of channel increases or the device footprint changes, it may be necessary to reselect the appropriate initial structure and algorithm, which will significantly increase the optimization time. Besides, fabrication imperfections can seriously affect theses device properties considering the structure features are required in the scale of nanometers.

Here, we propose a compact topological power splitter (TPS) based on a topological photonic crystal (TPC). By combining two TPCs with different topological phases, kink states with unidirectional propagation along the domain wall of their interface can be formed, which offers a novel approach to optical transmission with immunity to backscattering and fabrication defects [12-24]. The propagation mode in the kink states of TPC has a well-defined single mode over a large bandwidth relying on the linear dispersion relations. Therefore, the kink states can effectively suppress backscattering and thus allow ultra-low loss propagation through sharp turns with the improved uniformity over a wide wavelength range [25-27]. As shown in the left-top schematic structure of Fig. 1(a), a 1×2 TPS is designed by simply arranging valley photonic crystals (VPCs) with different topological phases (red and blue regions) in a splitter configuration. The distribution of the corresponding guiding mode ($y$ component of the electric field) through the TPS is shown in the right-top of Fig. 1(a), presenting well-confined mode at the interfaces of the VPCs with different phases. Benefiting from the valley-locked propagation mode of the VPCs, the two boundary states are unable to couple to each other in the same direction. As a result, the light could robustly walk around the sharp bending with little scattering, which also promises the compact footprint.

For comparison, the structures and mode distributions of power splitters based on conventional Y-junction and MMI are displayed in Figs. 1(b) and 1(c). In the Y-junction structure, parameters should be optimized to reduce the insertion loss in the Y-branch domain due to the mode mismatch. Besides, the perturbation of bending angle would also affect transmission efficiency, which has a tradeoff with the device footprint. The MMI structure would suffer from the same problem to fulfill the mode-beating condition. These two structures are also sensitive to disturbances caused by waveguide widths. These considerations complicate the design and fabrication of the power splitters remarkably.

In contrast, the proposed TPS is realized by simply configuring TPCs without the sophisticated design of splitting domain. The immunity of back-scattering in TPCs could eliminate the excess loss at the turns. In addition, the operation wavelength range is determined by the photonic band of the TPCs, promising a uniform power splitting over a large bandwidth. Besides, TPCs have more freedom of design by changing the primitive cell size to quickly adjust the interested wavelength ranges. Finally, the TPC promises the TPSs with multiple output ports in more compact footprints benefiting from the small angle and size at each bending.

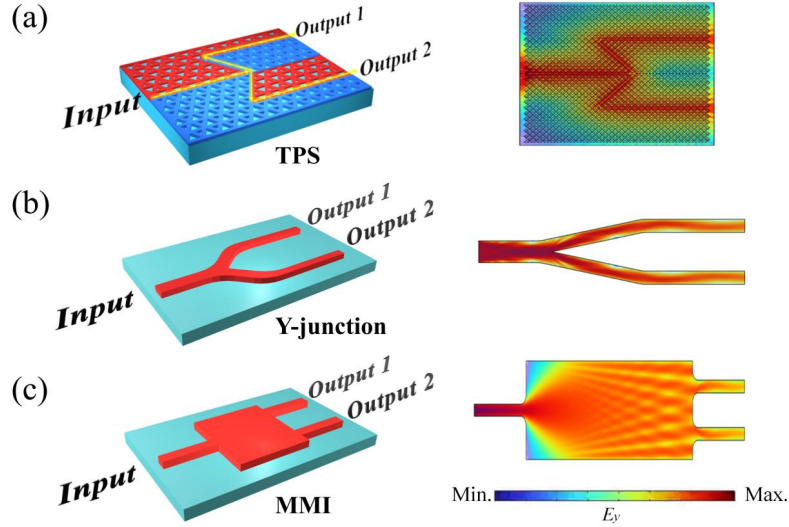

**Fig. 1.** Schematic structures (left) and calculated electric field $E_y$ profiles (right) of the 1×2 power splitters based on TPC (a), Y-junction (b), MMI (c). .

    The proposed TPS based on TPC is fabricated on a silicon-on-insulator platform with a $h$=220 nm thick top silicon slab. With the techniques of electron beam lithography and inductively coupled plasma etching, the TPS is defined in the top silicon slab. The design of the TPCs utilizes a honeycomb lattice of triangular air-holes with a lattice constant $a$=455 nm as the basic structure [20,23]. Figure 2(a) displays the scanning electron microscope (SEM) image of the fabricated TPC. The unit cell is shown in the red dashed lines, containing two reversed equilateral triangular holes with side lengths of $d_1$ and $d_2$ respectively. To illustrate the attributes of the TPCs, we implement the numerical calculations of its photonic band structure using the finite element method, as shown in Fig. 2(b). The air-hole type photonic crystal slab is more amenable to produce TE-mode photonic bandgaps, so we focus on the electric field confined in the slab. When the sides of the triangular air-holes are equal (for example, $d_1$= $d_2$= 240 nm), the TPCs exhibit inversion symmetry and $C_{6v}$ symmetry, which generates degenerate Dirac point at point K (K') in Brillouin zone (green dashed line in Fig. 2(b)) [16,17,23]. We can lift the degeneracy of the K (K') point by setting different $d_1$ and $d_2$ to obtain topologically valley photonic crystals (VPCs), which generates TE-mode topological photonic bandgaps. Here, we define $\Delta d = d_1-d_2$ to differentiate two spatially-reversed VPCs (VPC1($\Delta d$ >0), VPC2 ($\Delta d$ >0)). By setting the triangle side lengths with different values of 190 nm and 340 nm, due to mirror symmetry, the two VPCs have exactly same bands and produce a topological band gap from 187 THz to 205 THz, as depicted by the red lines and the yellow region in Fig. 2(b)

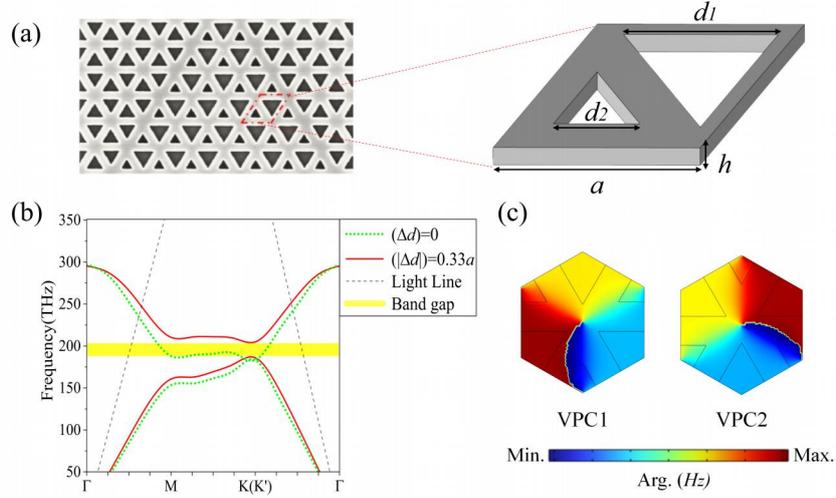

**Fig. 2.** (a) SEM image of the fabricated VPCs and the corresponding unit cell. (b) Band diagram of the VPC structures with different $\Delta d$. (c) Phase distribution of the magnetic field in the K-valley for the first band of VPC1 and VPC2.

As the two structures of VPC1 and VPC2 retain time-reversal symmetry, the Berry curvature integrals over the Brillouin zone is zero. However, at the extreme point of the energy band (K and K' valley), the local Berry curvature integrals are nonzero, so we define this value as a topological invariant, called the valley Chern number [16]. Numerical calculations indicate that VPC1 and VPC2 have opposite valley Chern numbers $C_{k/k'}=\pm 1/2$ at points K and K', indicating that they have different topological phases. The phase distributions of the magnetic field for VPC1 and VPC2 at point K are depicted in Fig. 2(c). When the modes of K valley are activated, the VPC1 has a left-handed circularly polarized phase vortex while the VPC2 possesses a right-handed circularly polarized phase vortex. The above distribution of Berry curvature and magnetic field phases at K (K') valley exhibits topological properties of the lattice primitive cells, and this topological valley chiral-locked polar distribution drives the valley Hall effect [16].

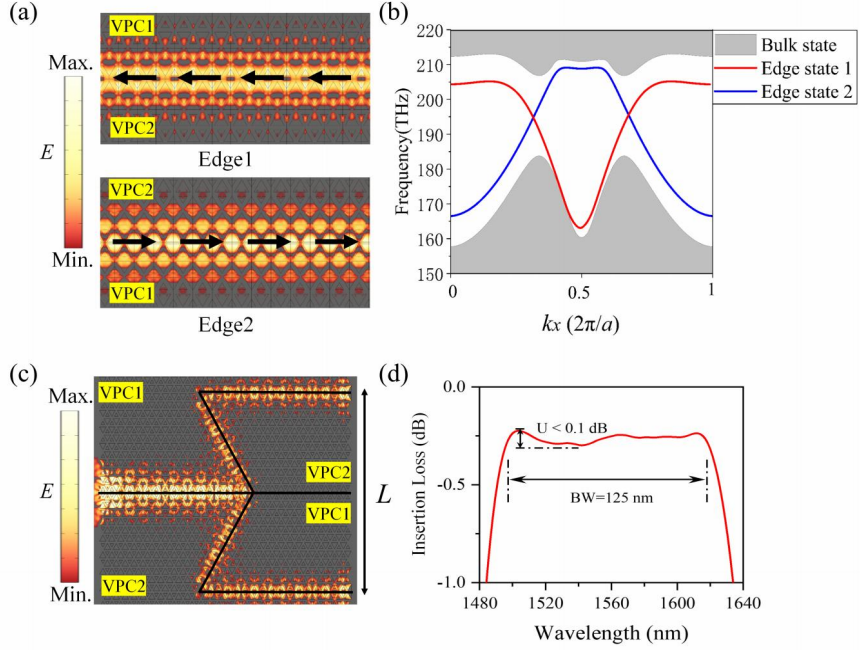

**Fig. 3.** (a) Propagation modes of the two different edges consisting of VPC1 and VPC2, showing the electric field distribution in K valley ($k_x$=0.3), where the black arrows indicate the direction of the Poynting vectors. (b) Dispersion of two edge states at K and K' valley. (c) Schematic of the 1×2 TPS and its electric field profiles distribution. (d) Insertion Loss (IL) of the proposed 1×2 TPS.

The basic optical channel of the proposed TPS is constructed by a kink-type domain wall between the VPC1 and VPC2. According to the bulk-boundary correspondence, a valley-polarized topological kink state that locked to the K (K') valley would appear at the domain wall and propagate forward within the bandgap [12,16,17]. VPC1 andVPC2 can compose two optical channels along their domain walls (topological edges), as shown in Fig. 3(a), and the corresponding dispersions of the valley kink states are depicted in Fig. 3(b). Each edge state has constant and opposite group velocities in the K and K' valleys, which contributes to the single directional transmission and immune backscattering properties of the topological modes [26,27]. The distributions of the Poynting vectors for both channels are shown by the black arrows in Fig. 3(a). It is clear to see that two topological kink states have opposite group velocity in the K valley ($k_x$=0.3), therefore they cannot couple to each other.

By composing VPC1 and VPC2, a 1×2 TPS can be formed as depicted in Fig. 3(c), where Edge1 and Edge2 compose a beam-splitting branch with a 120° turn angle. The spacing length between the two branches is defined as $L$. The light is incident from the left port and locked by the VPCs on both sides to transmit along the domain walls. Since the inter-valley scattering is suppressed and the two boundary states are unable to couple to each other in the same direction, the light could robustly walk around the sharp bendings with little scattering [25,27]. The two channels have exactly same valley topological modes, and therefore the light is able to achieve average power splitting at the branchs and ultimately low-loss output at both output ports.

According to the transmission characteristics of the topological interface of TPCs, there is no extra loss produced by changing the period of the TPCs while ensuring the existence of a topological bandgap [16]. Therefore, compact devices of arbitrary size can be fabricated under the design of sharp corners. Additionally, all of frequencies within the bandgap can form topological modes to transmit along the topological edges, so the TPC-based TPS has a large bandwidth (BW) and weak wavelength-dependence. A flat-top high transmission can be obtained within the bandgap and the light could be equally split with little splitting-ratio variation. These excellent characteristics will enable the designed TPS with good uniformity, low insertion loss and broad bandwidth. To verify the above expectations, the transmission of the proposed 1×2 TPS is simulated as well. As shown in Fig. 3(d), it can be obtained that TPS displays a good uniformity (U) of less than 0.1 dB and low insertion loss (IL<0.3 dB) with a flat-top transmission ranging from 1495 to 1620 nm.

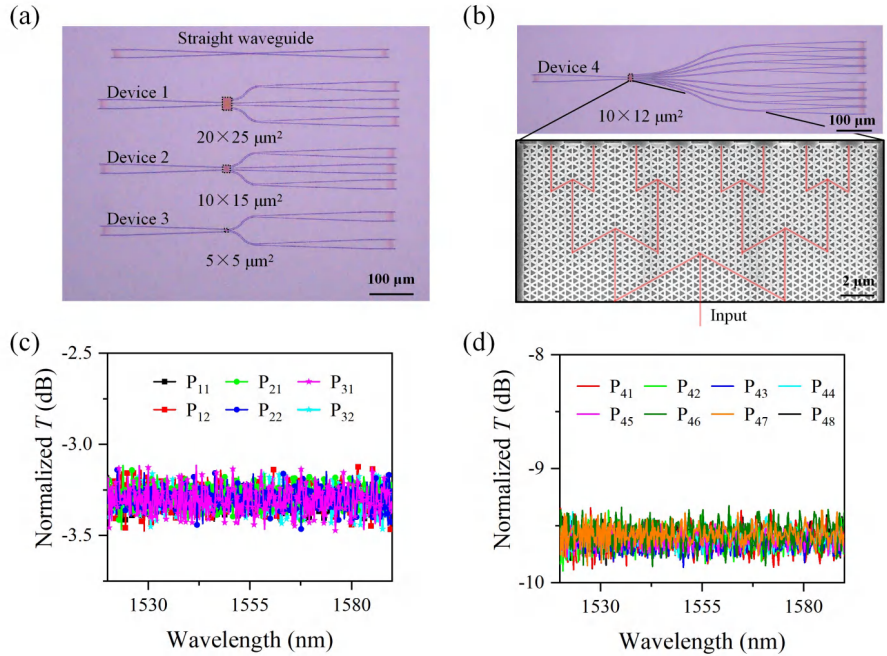

**Fig. 4.** (a) Optical micrograph of the fabricated 1×2 TPSs, which have different spacing lengths $L$ between the two branches of $3\sqrt{3}a$ (Device 1), $12\sqrt{3}a$ (Device 2) and $20\sqrt{3}a$ (Device 3). (b) Optical micrograph (top) and SEM image (bottom) of the fabricated 1×8 TPS. (c) Transmission spectra of the fabricated 1×2 TPSs with different $L$. (d) Transmission spectra of the fabricated 1×8 TPS.

To further present the reliability of the device, we fabricate and characterize a series of the proposed TPS. Figure 4(a) shows the optical micrograph of the fabricated 1×2 TPSs with different spacing length $L$ of $3\sqrt{3}a$ (Device 1), $12\sqrt{3}a$ (Device 2) and $20\sqrt{3}a$ (Device 3), respectively. To calibrate the TPSs, a straight TPC waveguide is fabricated as well. Grating couplers are designed at both ends of the waveguide to assist the light coupling-in and -out with the optical fibers. These devices are characterized by coupling a narrowband tunable laser into the input grating coupler. The transmission powers are monitored by an optical power meter at the other grating coupler and normalized by the TPC waveguide. Measured normalized transmissions of all three devices are displayed in Fig. 4(c).

Although the spacing lengths $L$ between the two output ports are decreased among Device 1, Device 1, and Device 3, the crosstalk does not increase observably due to the topologically-protected transport. All devices exhibit good transmission uniformity of less than 0.35 dB with a broad bandwidth ranging from 1520 to 1590 nm. The bandwidth (70 nm) is smaller than the simulated results (125 nm), which is limited by the bandwidth of the input and output grating couplers [17]. The maximum insertion loss of 0.38 dB within the bandwidth is obtained. Besides, the total footprint of the 1×2 TPS is reduced to a compact size of only 5×5 $\mu m^2$.

Table 1. Comparison of the 1×8 TPS based on TPC and conventional structures.

| Refs. | Structures | Method | Channel | IL/dB | U/dB | BW/nm | Footprint/$\mu m^2$ |
|---|---|---|---|---|---|---|---|
| [6] | Y-junction | experimental | 8 | 1.28 | 0.11 | 100 | >1000×1000 |
| [8] | PC | numerical | 8 | <0.45 | - | 55 | 341.9 |
| [3] | MMI | experimental | 8 | 0.65 | 0.84 | 104 | 11.3×47.8 |
| This work | TPC | experimental | 8 | 0.58 | 0.6 | 70 (125) | 10×12 |

Based on this rule, a compact 1×8 TPS is demonstrated as well. The optical micrograph of the fabricated device is shown in top image of Fig. 4(b). The bottom image of Fig. 4(b) gives the enlarged SEM image of the TPS region with a compact size of 10×12 $\mu m^2$. Corresponding normalized transmissions from the 8 output ports are plotted in Fig. 4(d) and all exhibit good uniformity (U<0.6 dB) within the bandwidth of 70 nm. This larger U compared with that of the 1×2 TPS is mainly owing to the Fabry-Perot interference between the interface caused ripples, while showing a weak wavelength dependence. To evaluate the performance of the fabricated 1×8 TPS, we list the detailed performances of power splitters reported in literatures based on other structures. Compared with these works, our device shows a good uniformity and low insertion loss over a broad bandwidth in the most compact footprint.

In conclusion, we have proposed and demonstrated on-chip 1×N power splitters based on the topologically protected interface channel of TPCs. Benefiting from the valley-locked propagation mode of the valley-type TPCs, the proposed TPS offers an optical transmission with immunity to fabrication defects and backscattering around the sharp bendings. It therefore enables the low insertion loss, good uniformity, and compact footprint. For the fabricated 1×2 (8) power splitter, the uniformity is below 0.35 (0.65) dB and the corresponding maximum insertion loss is 0.38 (0.58) dB with compact footprint of 5×5 $\mu m^2$ (10×12 $\mu m^2$) within a bandwidth from 1520 to 1590 nm. Compared with the power splitters based on conventional structures, the 1×N TPS enabled by the topological photonics maintains high device performance in more compact footprint, which is more appropriate for the optical processing in silicon PICs. In addition, the TPS only requires simple configurations of VPCs with different topological phases, which is easier and more reliable for designs and fabrications.

**Funding.** The Key Research and Development Program (2022YFA1404800), National Natural Science Foundation of China (12374359, 62375225, 62305270), Shaanxi Fundamental Science Research Project for Mathematics and Physics (22JSY004), Xi'an Science and Technology Plan Project (2023JH-ZCGJ-0023) .

**Acknowledgments.** The authors would thank the Analytical & Testing Center of NPU for the assistances of device fabrication.

**Disclosures.** The authors declare no conflicts of interest.

**Data availability.** Data underlying the results presented in this paper are not publicly available at this time but may be obtained from the authors upon reasonable request.

**References**

1. H. Yamada, C. Tao, S. Ishida, et al., "Optical directional coupler based on Si-wire waveguides," IEEE Photonics Technology Letters **17**, 585-587 (2005).
2. I. Park, H.-S. Lee, H.-J. Kim, et al., "Photonic crystal power-splitter based on directional coupling," Optics Express **12**(2004).
3. R. Yao, H. Li, B. Zhang, et al., "Compact and Low-Insertion-Loss 1×N Power Splitter in Silicon Photonics," Journal of Lightwave Technology **39**, 6253-6259 (2021).
4. D. J. Thomson, Y. Hu, G. T. Reed, et al., "Low Loss MMI Couplers for High Performance MZI Modulators," IEEE Photonics Technology Letters **22**, 1485-1487 (2010).
5. Z. Han and S. He, "Multimode interference effect in plasmonic subwavelength waveguides and an ultra-compact power splitter," Optics Communications **278**, 199-203 (2007).
6. S. H. Tao, Q. Fang, J. F. Song, et al., "Cascade wide-angle Y-junction 1 × 16 optical power splitter based on silicon wire waveguides on silicon-on-insulator," Optics Express **16**(2008).
7. Y. Wang, S. Gao, K. Wang, et al., "Ultra-broadband and low-loss 3 dB optical power splitter based on adiabatic tapered silicon waveguides," Optics Letters **41**(2016).
8. I. Moumeni and A. Labbani, "Very high efficient of 1 × 2, 1 × 4 and 1 × 8 Y beam splitters based on photonic crystal ring slot cavity," Optical and Quantum Electronics **53**(2021).
9. T. Din Chai, T. Kambayashi, S. R. Sandoghchi, et al., "Efficient, Wide Angle, Structure Tuned 1×3 Photonic Crystal Power Splitter at 1550 nm for Triple Play Applications," Journal of Lightwave Technology **30**, 2818-2823 (2012).
10. K. Xu, L. Liu, X. Wen, et al., "Integrated photonic power divider with arbitrary power ratios," Optics Letters **42**(2017).
11. T. Tian, Y. Liao, X. Feng, et al., "Metasurface‐Based Free‐Space Multi‐Port Beam Splitter with Arbitrary Power Ratio," Advanced Optical Materials **11**(2023).
12. X.-T. He, E.-T. Liang, J.-J. Yuan, et al., "A silicon-on-insulator slab for topological valley transport," Nature Communications **10**(2019).
13. M. I. Shalaev, W. Walasik, A. Tsukernik, et al., "Robust topologically protected transport in photonic crystals at telecommunication wavelengths," Nature Nanotechnology **14**, 31-34 (2018).
14. F. Gao, H. Xue, Z. Yang, et al., "Topologically protected refraction of robust kink states in valley photonic crystals," Nature Physics **14**, 140-144 (2017).
15. L.-H. Wu and X. Hu, "Scheme for Achieving a Topological Photonic Crystal by Using Dielectric Material," Physical Review Letters **114**(2015).
16. T. Ma and G. Shvets, "All-Si valley-Hall photonic topological insulator," New Journal of Physics **18**(2016).
17. L. Gu, Q. Yuan, Q. Zhao, et al., "A Topological Photonic Ring-Resonator for On-Chip Channel Filters," Journal of Lightwave Technology **39**, 5069-5073 (2021).
18. Q. Yuan, L. Gu, L. Fang, et al., "Giant Enhancement of Nonlinear Harmonic Generation in a Silicon Topological Photonic Crystal Nanocavity Chain," Laser & Photonics Reviews **16**(2022).
19. H. Wang, S. K. Gupta, B. Xie, et al., "Topological photonic crystals: a review," Front Optoelectron **13**, 50-72 (2020).
20. Y. Yang, Y. Yamagami, X. Yu, et al., "Terahertz topological photonics for on-chip communication," Nature Photonics **14**, 446-451 (2020).
21. L. Gu, B. Wang, Q. Yuan, et al., "Fano resonance from a one-dimensional topological photonic crystal," APL Photonics **6**(2021).
22. A. Kumar, Y. J. Tan, N. Navaratna, et al., "Slow light topological photonics with counter-propagating waves and its active control on a chip," Nat Commun **15**, 926 (2024).
23. H. Wang, L. Sun, Y. He, et al., "Asymmetric Topological Valley Edge States on Silicon-On-Insulator Platform," Laser & Photonics Reviews **16**(2022).
24. M. I. Shalaev, W. Walasik, A. Tsukernik, et al., "Robust topologically protected transport in photonic crystals at telecommunication wavelengths," Nat Nanotechnol **14**, 31-34 (2019).
25. A. B. Khanikaev and A. Alu, "Topological photonics: robustness and beyond," Nat Commun **15**, 931 (2024).
26. Y. Yang, Y. F. Xu, T. Xu, et al., "Visualization of a Unidirectional Electromagnetic Waveguide Using Topological Photonic Crystals Made of Dielectric Materials," Phys Rev Lett **120**, 217401 (2018).
27. M. C. Rechtsman, "Reciprocal topological photonic crystals allow backscattering," Nature Photonics **17**, 383-384 (2023)